\documentclass{openjournal}

\usepackage{xcolor}
\usepackage{textgreek}
\usepackage[utf8]{inputenc}
\usepackage[english]{babel}
\usepackage{multirow}
\usepackage{arydshln}
\usepackage{booktabs}   

\usepackage[percent]{overpic}
\usepackage{graphicx}
\usepackage{txfonts}
\usepackage{orcidlink}
\usepackage{graphicx} 
\usepackage{float}    
\usepackage{hyperref}
\hypersetup{
    unicode, 
    colorlinks=true,
    linkcolor=linkcolor,
    citecolor=linkcolor,
    filecolor=linkcolor,
    urlcolor=linkcolor,
}
\usepackage{color,colortbl}
\definecolor{linkcolor}{rgb}{0.0,0.3,0.5}
\usepackage{tensind}
\tensordelimiter{?}
\usepackage{natbib}

\DeclareGraphicsExtensions{.png,.jpg,.pdf}
\usepackage{verbatim}
\usepackage[normalem]{ulem}
\usepackage{orcidlink}
\usepackage{soul}

\graphicspath{ {./figs/} }

\begin{document}

\title{Orbital stability of moons around the TRAPPIST-1 planets}

\author{Shubham Dey\orcidlink{0009-0006-0865-8562}}
\email{ssdey99@gmail.com}
\affiliation{Indian Institute of Science Education Research, Kolkata, India}

\author{Sean N. Raymond\orcidlink{0000-0001-8974-0758}}
\email{rayray.sean@gmail.com}
\affiliation{Laboratoire d’Astrophysique de Bordeaux, CNRS and Université de Bordeaux, Allée Geoffroy St. Hilaire, 33165 Pessac, France}

\begin{abstract}
    
    We investigate the dynamical stability of potential satellites orbiting the seven planets of the \texttt{TRAPPIST-1} system using a suite of $N$-body simulations. For each planet, we show that moons can remain stable from the Roche limit out to near the theoretical prograde stability boundary at roughly $0.5$ Hill Radii. We quantify how perturbations from neighbouring planets modify these stability limits. Although the overall effect of individual perturbers is generally weak, the combined gravitational influence of the full multi-planet configuration produces a modest contraction of the outer stable radius, notably for \texttt{TRAPPIST-1 b} and \texttt{TRAPPIST-1 e}. For each of the seven planets, the outer stability limit for satellites is at 40-45\% of the Hill radius, consistent with previous work.  Using simple long-term tidal decay calculations, we show that the most massive satellites that could survive over Gyr timescales are $10^{-(7-9)} M_\oplus$ (with higher possible masses for the outer planets). 
\end{abstract}


\maketitle

\section{Introduction}

Previous studies have investigated the stability of satellites orbiting a planet using both analytical \citep{szebehely1978stability, graziani1981orbital, pendleton1983further, hamilton1991orbital, donnison2010hill} and numerical \citep{holman1999long, domingos2006stable, payne2013stability, quarles2020orbital} techniques. They reveal that prograde satellite orbits are stable up to a maximum semi-major axis of approximately $0.5 \, R_H$, where $R_H$ represents the planet's Hill radius, $R_H = a (m/3 M_\star)^{1/3}$, where $a$ is the planet's orbital semimajor axis and $m$ its mass, and $M_\star$ is the stellar mass. More precisely, the stability limit for prograde moons $a_E$ can be expressed as ~\citep{domingos2006stable}:
\[
a_E \approx 0.4895 \ (1.0000 - 1.0305\ e_{\text{P}} - 0.2738\ e_{\text{sat}})\ \text{R}_{\text{H}},
\]
\noindent where $e_{\text{P}}$ and $e_{\text{sat}}$ are the planet's and satellite's eccentricities, respectively.  There is a corresponding relation for retrograde satellites that extends out to approximately the full edge of the planet's Hill radius.  Highly inclined orbits (around $90^\circ$) are generally unstable, and giant planet–planet scattering has been found to significantly disrupt satellite systems~\citep{gong2013effect,hong2018innocent}. In systems with many tightly-packed planets, the stable region for moons may be truncated, with the outer region perturbed by the gravitational influence of other planets~\citep{payne2013stability}.

Here, we dynamically test the stability of moons orbiting the planets of the \texttt{TRAPPIST-1} system.  This iconic system contains seven roughly Earth-sized planets in a multi-resonant orbital configuration orbiting an ultracool dwarf star \citep{gillon2017seven, luger2017seven, grimm2018nature, agol2021refining}. The planets' masses have been well characterized using transit-timing variations~\citep{agol2021refining}, enabling in-depth study of the origin of the resonant configuration~\citep{pichierri2024formation} and constraining the planets' bombardment histories~\citep{raymond2022upper}.  Given its very compact configuration, \texttt{TRAPPIST-1} is an enticing system in which to test the stability of moons.



The structure of this paper is organized as follows: In Section~\ref{sec:Simulations}, we outline the configuration of the system and present the benchmarking of the outer stability region, comparing our results with the findings of \citet{domingos2006stable}. It also examines the influence of neighboring planets, highlighting their perturbative effects on the orbital radius of satellites. Finally, in Section~\ref{sec:Discussion}, we summarize our findings, discuss the implications of this study, and emphasize the significance of our results. Additionally, we explore how external perturbations further alter the stability radius.

\section{Simulations}\label{sec:Simulations}

We directly simulated the orbital stability of moons of the \texttt{TRAPPIST-1} planets.The analysis employed the best-fit orbital architecture reported by \citet{agol2021refining}, with parameter values taken from Extended Data Table 2 of \citet{raymond2022upper}. These values are provided again in Appendix~\ref{sec:APPENDIX} for completeness. The central star's mass was fixed at $0.09 \rm{M_\star}$, and massless moons were placed on coplanar, initially circular orbits around one or more of the planets. For our simulations, we used the 15th-order Integrator with Adaptive Step-size control (\texttt{IAS-15}, \citep{carusi1985one}, \citep{rein2015ias15}) integrator within the \texttt{REBOUND} \citep{rein2012rebound} N-body simulation package.  We used a simulation timestep of $\sim 5\%$ of the planet's orbital period and simulate the system for $10^6$ times the planet's orbital period.  


As a first step, we tested the stability of moons around each Trappist-1 planet in isolation -- that is, without any other planets in the system.  We started by distributing 100 moons on initially circular, prograde orbits, extending from the planet's Roche limit to half of its Hill radius.  Next, we included the other planets in the system, one at a time, to determine their effect on moon stability. The analysis focused on identifying whether the presence of each neighboring body led to earlier ejection, collision events, or significant orbital perturbations of the moons.

\begin{figure*}[p]
    \centering
    \includegraphics[width=0.95 \textwidth]{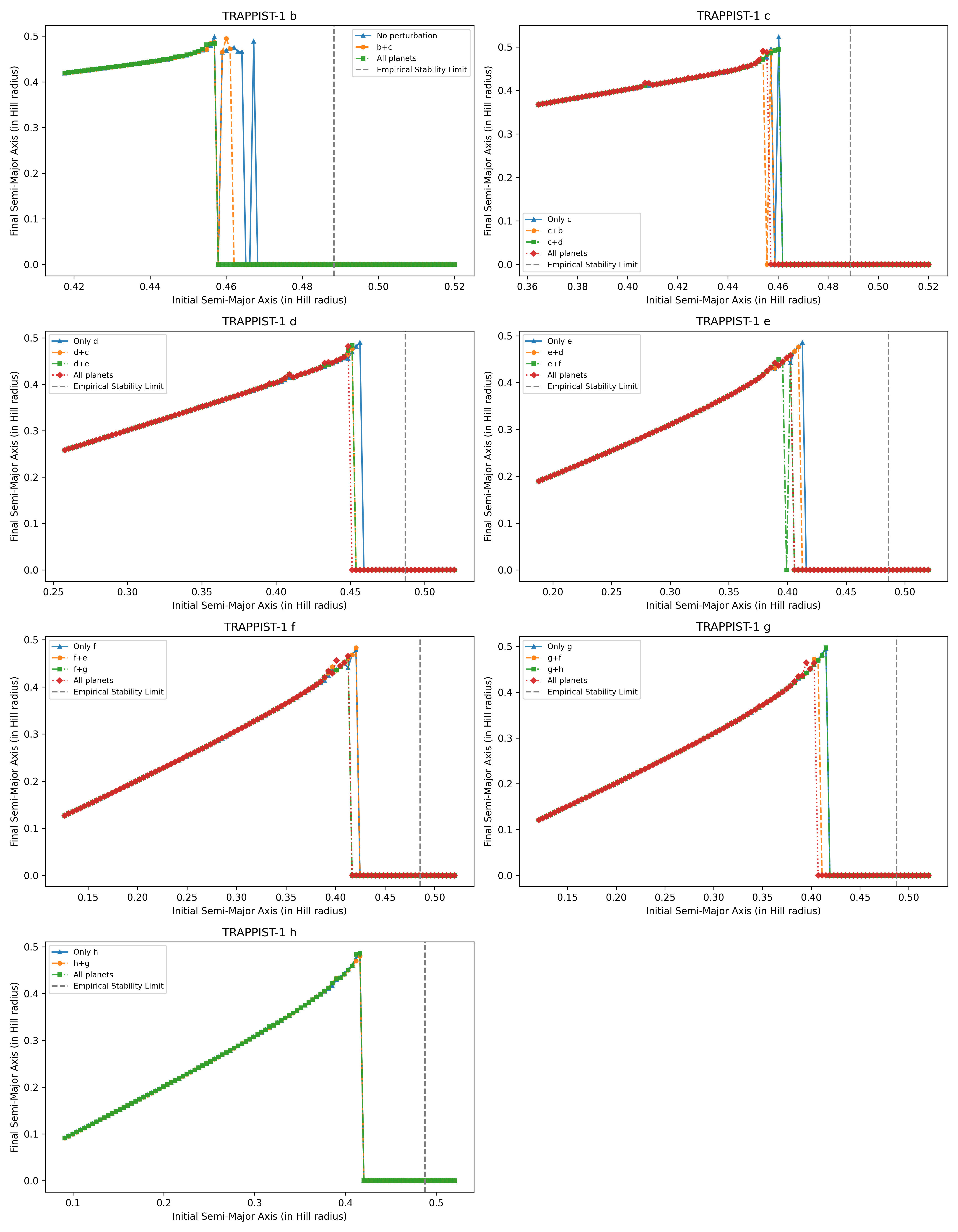}
    \caption{Final semi-major axes of satellites versus their initial orbital distance (in Hill radii) for each TRAPPIST-1 planet. Each panel compares stability outcomes under different perturbations: none, individual neighbouring planets, and the full multi-planet system. A drop to zero marks unstable orbits. The vertical dashed line indicates the empirical stability limit. In all cases, the full system shifts the stability boundary inward compared to isolated or single-perturber configurations.}
    \label{fig:summary}
\end{figure*}

The overall outer stability limit for each planet ranged from roughly 0.4 to 0.45 Hill radii.  This is slightly smaller than the analytical criterion from \cite{domingos2006stable}, but it is roughly consistent with the stability limits found in simulations of dynamically-evolving planetary systems~\citep[e.g.][]{payne2013stability,bolmont2025survival}. A comparison of the resulting outer stability radii revealed that, although most planets exert only modest dynamical effects, two cases --- \texttt{TRAPPIST-1 b} and \texttt{TRAPPIST-1 e} --- exhibited notable reductions in the size of their stable satellite regions, typically on the order of a percent of the Hill radius. The resulting stability limits are illustrated in Fig.~\ref{fig:summary}, and the key cases are discussed below.

For \texttt{TRAPPIST-1 b}, the only immediate neighbour is \texttt{TRAPPIST-1 c}. To evaluate its influence, three configurations were examined: (i) the planet considered in isolation, (ii) the planet together with \texttt{TRAPPIST-1 c}, and (iii) the planet embedded within the full TRAPPIST-1 system. As illustrated in Fig.~\ref{fig:summary}, including all neighbouring planets results in the greatest contraction of the outer stability radius, reducing it from $0.467\,R_{\rm H}$ (isolated case) to $0.457\,R_{\rm H}$ (that is, the stability limit shrank by 1\% of the Hill radius). Accounting only for \texttt{TRAPPIST-1 c} produces a smaller reduction, yielding a stability boundary of $0.461\,R_{\rm H}$. These results indicate that the overall destabilisation is not dominated by a single perturber but rather arises from the combined gravitational influence of the full planetary system.

For \texttt{TRAPPIST-1 e}, which is flanked by \texttt{TRAPPIST-1 d} and \texttt{TRAPPIST-1 f}, four scenarios were considered: the isolated system; the system including \texttt{TRAPPIST-1 d}; the system including \texttt{TRAPPIST-1 f}; and the full multi-planet configuration. The simulations reveal that \texttt{TRAPPIST-1 f} produces the stronger destabilising effect, reducing the outer stability boundary from $0.412\,R_{\rm H}$ (isolated case) to $0.403\,R_{\rm H}$. In comparison, \texttt{TRAPPIST-1 d} yields a more modest reduction to $0.409\,R_{\rm H}$. When all planets are included, the stability boundary decreases even further than in the presence of \texttt{TRAPPIST-1 d} alone, demonstrating that cumulative gravitational effects can surpass those of any individual neighbour.

\section{Discussion}\label{sec:Discussion}

The stability analysis performed for each planet in the \texttt{TRAPPIST-1} system shows that the combined gravitational influence of all neighbouring planets produces the most significant contraction of the outer stable radius of any potential satellite system (see Fig.~\ref{fig:summary}). This behaviour is consistently observed across the full set of POIs, indicating that multi-planet perturbations play a dominant role in shaping the extent of stable circumplanetary regions. However, although the fully perturbed configuration always yields the smallest stability boundary, the magnitude of this contraction is very small, and the influence of individual neighbouring planets is comparatively weak, at the level of 1\% of the Hill radius or less.


The simulations presented here neglect several additional physical processes that may further modify satellite stability over secular timescales, including tidal dissipation~\citep{tokadjian2020impact,patel2025tidally}, relativistic precession, mutual interactions between satellites, non-uniform gravitational fields and external perturbations \citep{kollmeier2019can}. For example, the moon has been found to have localized mass concentrations within its crust \citep{muller1968mascons} that destabilise the orbits of very close orbits around the Moon \citep{konopliv2001recent}.  Nevertheless, the simplified dynamical framework employed in this study provides meaningful constraints on the parameter space for viable exomoons in compact multi-planet systems. 

In this study, the maximum allowable masses of potential satellites orbiting the seven \texttt{TRAPPIST-1} planets were computed by applying a tidal stability constraint based on the formalism of \citet{barnes2002stability}. In this framework, a satellite remains long-lived only if its tidal orbital decay timescale exceeds the age of the system. Because the decay rate is controlled primarily by the tidal dissipation efficiency of the host planet, characterised by the ratio $k_2/Q$, the resulting stability limits depend sensitively on the adopted tidal parameters.


We adopt the analytic upper limit on the mass of a long-lived satellite, \(M_m\), obtained by requiring that the tidal migration timescale equals the star's age \(T\). The expression is~\citep{barnes2002stability}:

\begin{equation}
    M_m \leq 
    \frac{2}{13}
    \left(
        \frac{(f a_p)^3}{3 M_\ast}
    \right)^{13/6}
    \frac{
        M_p^{8/3} \, Q_p
    }{
        3 k_{2p} \, T \, R_p^5 \sqrt{G}
    } .
    \label{eq:moon_mass_limit}
\end{equation}

The parameters in Equation~\ref{eq:moon_mass_limit} are defined as follows:
In Equation~\ref{eq:moon_mass_limit}, 
$M_m$ is the maximum satellite mass that can survive tidal migration over a timescale $T$; 
$f$ denotes the fraction of the planetary Hill radius defining the outer stability boundary for prograde moons; 
$a_p$ is the planetary semi-major axis; 
$M_\ast$ is the host-star mass; 
$M_p$ and $R_p$ are the planetary mass and radius, respectively; 
$Q_p$ is the planetary tidal quality factor; 
$k_{2p}$ is the planet’s Love number; 
$T$ is the assumed system age; 
and $G$ is the gravitational constant.

This relation provides a conservative upper bound on the mass of any long-lived exomoon, assuming that tidal evolution dominates its orbital migration.\footnote{We note that several other studies have used more complex formalisms to explore the evolution of tidally-interacting planet-moon systems~\citep[e.g.][]{tokadjian2020impact,quarles2020orbital,patel2025tidally}.} To assess the impact of planetary tidal dissipation, we assume the tidal Love numbers $k_{2p}$ = 0.25 \citep{moore2000tidal}, and tidal quality factor $Q = 100$ \citep{lainey2016quantification}. We adopted 7~Gyr as the age of the star~\citep{burgasser2017age,gonzales2019reanalysis}. 

The maximum moon mass allowed from this tidal evolution calculation is extremely small, ranging from $\sim 10^{-10} M_\oplus$ for planet b to a few $\times 10^{-7} M_\oplus$ for planet h (see Table~\ref{tab:moonmasses}).  This corresponds to moons of diameter 80 to 280 km, assuming an asteroid-like density of 2.5 g/cm$^3$.  Clearly, these values are subject to large uncertainties in the planets' physical properties (e.g., their tidal quality factors), yet they are orders of magnitude below the masses of even the smallest regular moons in the Solar System. 

Taken together, these results suggest that, under tidal parameters consistent with rocky planets, the current \texttt{TRAPPIST-1} system is not expected to support a significant population of long-lived massive exomoons (for simulations of the tidal and dynamical evolution of TRAPPIST-1, see \citep{turbet2018modeling,bolmont2020solid,bolmont2020impact,
brasser2022long}. If any moons are present today, they are likely to be very small and exert only minimal dynamical influence.

\begin{table*}[t]
\centering
\renewcommand{\arraystretch}{1.25}
\caption{Maximum allowable satellite masses for the \texttt{TRAPPIST-1} planets assuming 
$k_{2,p}=0.25$, $Q=100$ ($k_2/Q = 2.5\times10^{-3}$), and a stellar age of 7 Gyr.}
\label{tab:moonmasses}

\begin{tabular}{|l|c|c|}
\hline
Planet &
\shortstack{Critical Radius \\ ($f = R_{\mathrm{crit}}/R_{\mathrm{Hill}}$)} &
\shortstack{Maximum Moon Mass \\ ($M_{\oplus}$)} \\
\hline

TRAPPIST-1 b & 0.457 & $7.60\times10^{-11}$ \\
TRAPPIST-1 c & 0.456 & $5.20\times10^{-10}$ \\
TRAPPIST-1 d & 0.449 & $9.67\times10^{-10}$ \\
TRAPPIST-1 e & 0.403 & $5.59\times10^{-9}$ \\
TRAPPIST-1 f & 0.413 & $6.75\times10^{-8}$ \\
TRAPPIST-1 g & 0.403 & $2.62\times10^{-7}$ \\
TRAPPIST-1 h & 0.416 & $3.20\times10^{-7}$ \\

\hline
\end{tabular}
\end{table*}

\section*{APPENDIX}\label{sec:APPENDIX}

\begin{table}[H]
    \centering
    
    \textbf{Orbital parameters of the TRAPPIST-1 system (taken from \citet{raymond2022upper})} \\[0.8em]

    \renewcommand{\arraystretch}{1.3} 
    \resizebox{\textwidth}{!}{%
    \begin{tabular}{|l|c|c|c|c|c|c|}
        \hline
        \textbf{Planet} & \textbf{Mass} ($M_\oplus$) & \textbf{Radius} ($R_\oplus$) & 
        \textbf{Axis} $a$ (AU) & \textbf{Eccentricity} $e$ & 
        \textbf{Longitude of periastron} $\varpi$ ($^\circ$) & 
        \textbf{Mean Anomaly} $\mathcal{M}$ ($^\circ$) \\
        \hline
        b & 1.3925 & 1.1174 & 0.011551 & 0.002344 & 253.61247 & 105.78489 \\
        c & 1.2943 & 1.0967 & 0.015820 & 0.001224 & 132.67293 & 54.89836 \\
        d & 0.3988 & 0.7880 & 0.022929 & 0.005922 & 49.34892 & 255.24260 \\
        e & 0.6924 & 0.9200 & 0.029300 & 0.002097 & 52.42927 & 30.97582 \\
        f & 1.0634 & 1.0448 & 0.038551 & 0.002898 & 170.04274 & 247.44087 \\
        g & 1.3464 & 1.1294 & 0.046896 & 0.003760 & 355.97174 & 87.27385 \\
        h & 0.3198 & 0.7552 & 0.061963 & 0.003571 & 172.18673 & 118.58431 \\
        \hline
    \end{tabular}
    } 

    \caption{Starting orbital configuration of the \texttt{TRAPPIST-1} system used in our fiducial simulations. 
    The parameters are adopted from \citet{raymond2022upper}. \\[1em]
    This table provides the initial dynamical state of the planetary system that is used to integrate the evolution of potential exomoons. 
    The parameters are derived from transit-timing variation analyses and are consistent with the long-term resonant architecture of the \texttt{TRAPPIST-1} planets.}
    \label{tab:trappist1_config}
\end{table}


\bibliographystyle{aa} 
\bibliography{oja_template} 

\end{document}